%% file: IEEE-conference-template-062824.tex
\documentclass[conference]{IEEEtran}
\IEEEoverridecommandlockouts

\usepackage{cite}
\usepackage{amsmath,amssymb,amsfonts}
\usepackage{algorithmic}
\usepackage{graphicx}
\usepackage{textcomp}
\usepackage{xcolor}
\usepackage{soul,color}
\usepackage{tabularx}
\usepackage{url}

\def\BibTeX{{\rm B\kern-.05em{\sc i\kern-.025em b}\kern-.08em
    T\kern-.1667em\lower.7ex\hbox{E}\kern-.125emX}}
\begin{document}

\title{IPsec based on Quantum Key Distribution: Adapting non-3GPP access to 5G Networks to the Quantum Era\\


}

\author{\IEEEauthorblockN{Asier Atutxa}
\IEEEauthorblockA{\textit{\textsuperscript{a}Dept. of Communications Engineering} \\
\textit{\textsuperscript{b}EHU Quantum Center} \\
\textit{University of the Basque Country}\\
Bilbao, Spain \\
asier.atutxa@ehu.eus}
\and
\IEEEauthorblockN{Ane Sanz}
\IEEEauthorblockA{\textit{\textsuperscript{a}Dept. of Communications Engineering} \\
\textit{\textsuperscript{b}EHU Quantum Center} \\
\textit{University of the Basque Country}\\
Bilbao, Spain \\
ane.sanz@ehu.eus}
\and
\IEEEauthorblockN{Eire Salegi}
\IEEEauthorblockA{\textit{\textsuperscript{a}Dept. of Communications Engineering} \\
\textit{University of the Basque Country}\\
Bilbao, Spain \\
eire.salegi@ehu.eus}
\and
\IEEEauthorblockN{Gaizka González}
\IEEEauthorblockA{\textit{\textsuperscript{a}Dept. of Communications Engineering} \\
\textit{University of the Basque Country}\\
Bilbao, Spain \\
ggonzalez088@ikasle.ehu.eus}
\and
\IEEEauthorblockN{Jasone Astorga}
\IEEEauthorblockA{\textit{\textsuperscript{a}Dept. of Communications Engineering} \\
\textit{\textsuperscript{b}EHU Quantum Center} \\
\textit{University of the Basque Country}\\
Bilbao, Spain \\
jasone.astorga@ehu.eus}
\and
\IEEEauthorblockN{Eduardo Jacob}
\IEEEauthorblockA{\textit{\textsuperscript{a}Dept. of Communications Engineering} \\
\textit{\textsuperscript{b}EHU Quantum Center} \\
\textit{University of the Basque Country}\\
Bilbao, Spain \\
eduardo.jacob@ehu.eus}
}

\maketitle

\begin{abstract}
The advent of quantum computing will pose great challenges to the current communication systems, requiring essential changes in the establishment of security associations in traditional architectures. In this context, the multi-technological and heterogeneous nature of 5G networks makes it a challenging scenario for the introduction of quantum communications. Specifically, 5G networks support the unification of non-3GPP access technologies (i.e. Wi-Fi), which are secured through the IPsec protocol suite and the Non-3GPP Interworking Function (N3IWF) entity. These mechanisms leverage traditional public key cryptography and Diffie-Hellman key exchange mechanisms, which should be updated to quantum-safe standards. Therefore, in this paper we present the design and development of a Quantum Key Distribution (QKD) based non-3GPP access mechanism for 5G networks, integrating QKD keys with IPsec tunnel establishment. Besides, we also demonstrate the feasibility of the system by experimental validation in a testbed with commercial QKD equipment and an open-source 5G core implementation. Results show that the time required to complete the authentication and IPsec security association establishment is 4.62\% faster than traditional cryptography PSK-based systems and 5.17\% faster than the certificate-based system, while ensuring Information-Theoretic Security (ITS) of the QKD systems.
\end{abstract}


\section{Introduction}


Fifth-generation mobile networks (5G) have become a critical part of current communication infrastructures. These networks support a variety of applications, including ultra-reliable and low-latency communications, as well as massive machine-type communications. From a security perspective, 5G relies on a combination of native 3GPP mechanisms and widely used transport and network security protocols that ensure confidentiality, authentication, and data integrity, such as TLS or IPsec~\cite{tang2022systematic}. These protocols currently build on public key primitives for authentication and key exchange, leveraging mechanisms such as Diffie-Hellman (D-H).

With the advent of quantum computing, traditional public-key cryptographic algorithms may render vulnerable due to the development of the Shor's algorithm~\cite{mavroeidis2018impact}. This potential paradigm shift in cryptography also presents risks to 5G networks. Consequently, advancing towards quantum-safe 5G deployments involves replacing or enforcing classical key establishment mechanisms with quantum-safe alternatives, such as Quantum Key Distribution (QKD).

In this context, the segments between access and core sites are specially relevant, as they often span geographically separated locations where maintaining security guarantees is essential. Current 5G deployments typically protect these interfaces using IPsec tunnels established via IKEv2. The security of these tunnels relies on the symmetric keys negotiated by IKEv2, which are derived from classical Diffie-Hellman key exchange mechanisms and are therefore vulnerable to quantum attacks.

\begin{figure}[h]
\centering
\includegraphics[width=1\columnwidth]{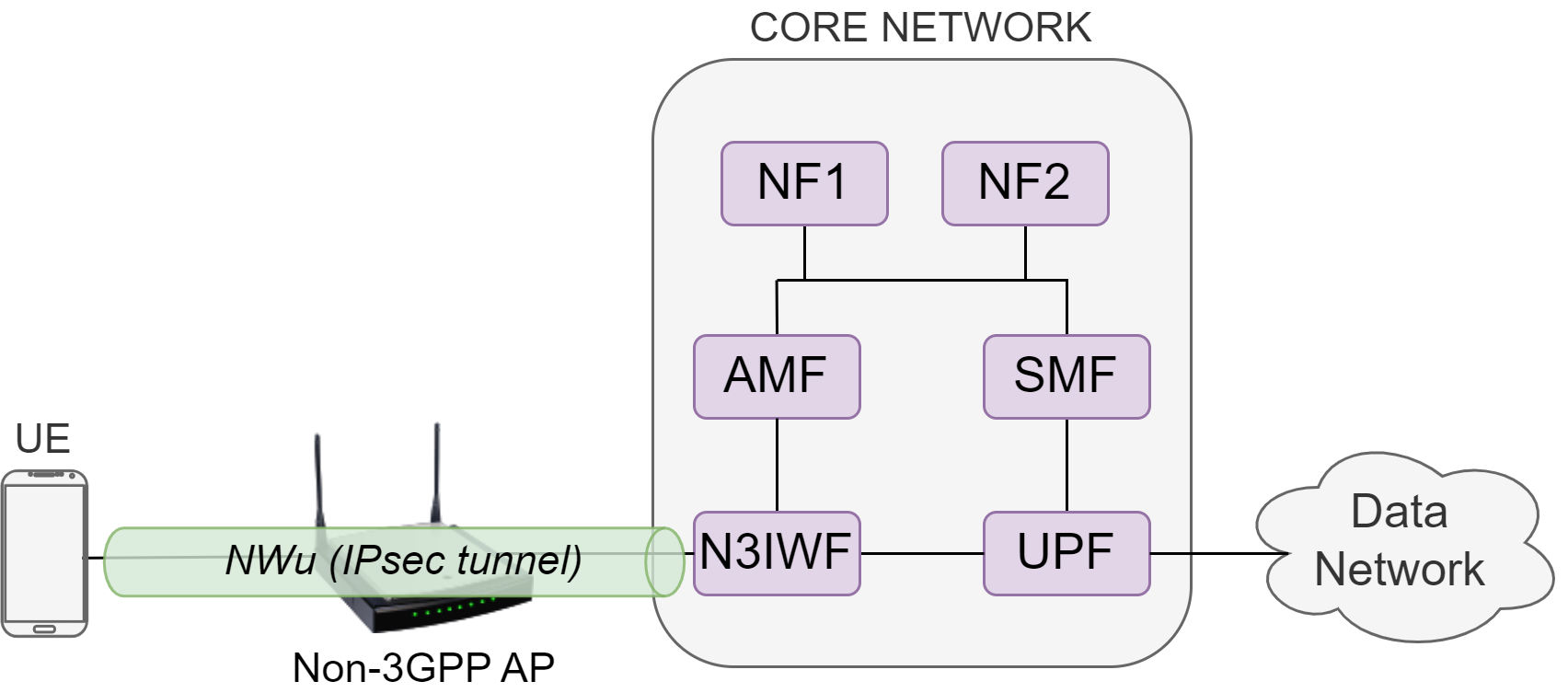}
\caption{Overview of the 5G network architecture for non-3GPP access.}
\label{fig:non-3gpp-access}
\end{figure}

Additionally, one of the key features introduced by 5G networks is the heterogeneity, bringing and integrating multi-technology systems within the same network. Thus, with the aim of incorporating additional networks beyond 3GPP radio access, 5G also supports non-3GPP access technologies, such as Wi-Fi, that are integrated into the 5G core through a specific interworking architecture, as depicted in Fig.~\ref{fig:non-3gpp-access}. In the untrusted non-3GPP case, the User Equipment (UE) reaches the 5G core via the Non-3GPP Interworking Function (N3IWF), and security between the UE and the N3IWF is provided by IPsec tunnels established over the NWu interface using IKEv2~\cite{kunz2020non}. These tunnels carry both control-plane and user-plane traffic, and are therefore exposed to the quantum vulnerabilities of the underlying IKEv2 key establishment. As a result, non-3GPP access constitutes a relevant scenario to integrate quantum-safe key establishment in 5G networks.

Therefore, this work proposes and experimentally evaluates a QKD-based non-3GPP access mechanism for 5G networks. More specifically, the main contributions of the work are:

\begin{enumerate}
    \item Integration of QKD keys into the tunnel establishment of untrusted non-3GPP access of 5G networks.
    \item Implementation of the proposed system in a hardware testbed combining commercial QKD equipment and an open-source 5G core implementation.
    \item Performance assessment of the proposed system compared to the standard certificate-based and Pre-Shared Key-based (PSK-based) systems.
\end{enumerate}

The rest of the paper is organized as follows. Section~\ref{sec:sota} presents a summary of the related work. Section~\ref{sec:proposed-system} describes the proposed QKD-based non-3GPP access mechanism for 5G networks. Then, Section~\ref{sec:testbed} describes the testbed used for the performance evaluation and the testing methodology, while Section~\ref{sec:results} overviews the obtained results. Finally, Section~\ref{sec:conclusions} provides a summary of the key findings and conclusions of the paper.

\section{Related Work}
\label{sec:sota}

The integration of quantum communications into the security of optical fronthaul and backbone networks has been a recent topic in the research community. Current 5G networks, as well as next generation networks (i.e. 6G), heavily rely on the heterogeneity and multi-technology, inevitably increasing the complexity of transitioning to security in the quantum era. 

Early work on 5G and beyond 5G networks focused on embedding QKD equipment in the fronthaul. In~\cite{zavitsanos}, Zavitsanos et al. propose to secure eCPRI transport layers using QKD for key exchanges between the centralized Baseband Unit (BBU) and distributed 5G terminal nodes. Similarly, in~\cite{zhang} Zhang et al. present an architecture that enables the coexistence of quantum and classical channel leveraging QKD with Wavelength Division Multiplexing (WDM) for 6G fronthaul networks. These coexistence solutions based on WDM systems can also encrypt the data channels at line rate using keys generated by the QKD equipment, as demonstrated by Sanz et al.~\cite{sanz}, achieving successful long-range encryption for fronthaul deployments.

Additional works focus on embedding QKD into the security protocols used in each network segment of 5G networks, such as TLS and IPsec. For instance, Atutxa et al.~\cite{atutxa2024towards} integrate and experimentally validate a QKD infrastructure in the 5G core, where TLS-protected service-based interfaces between network functions are supplied with QKD keys. In a complementary direction, Wright et al.~\cite{wright20215g} extend the 5G network slicing model by introducing a Software-Defined Networking (SDN)-based orchestrator capable of provisioning slices with different encryption mechanisms, including QKD-based encryption for slices with stringent security requirements. 

However, not many more works focus on adapting the security protocols in the context of 5G scenarios. Instead, most of the recent works aim at the link and network layer protocols rather than the 5G-specific transmission. For instance, preliminary work by Neppach et al.~\cite{neppach} presents a key management approach for integrating quantum-generated keys from QKD into the IPsec/IKE protocol framework, describing the design and implementation of a security gateway that combines QKD and IPsec within a single system-on-chip solution, addressing the mismatch between QKD's continuous key production and IPsec's periodic re-keying needs. Marksteiner~\cite{marksteiner} proposes a solution for the fast rekeying environment of IPsec tunnels through QKD-generated keys. Following this work, Dervisevic et al.~\cite{dervisevic} present an overview of the efforts of leveraging QKD generated keys in IPsec tunnel establishments instead of DH keys. IPSeQ~\cite{gao}, a security-enhanced IPsec variant, incorporates QKD with dynamic sliding-window re-keying to handle key scarcity, achieving up to 10 Gbps throughput in tests with commercial QKD devices and routers.

Even though several efforts have focused on embedding quantum technologies, and more specifically QKD, into 5G heterogeneous networks, integration with the security establishment using IPsec in communications through the non-3GPP access have not been studied. Precisely, the multi-technology nature of the 5G architecture and the objective of unifying the management of diverse radio access networks requires a thorough analysis of the non-3GPP access, which leverages the N3IWF and the NWu interface. To support this approach, securing this interface is crucial for fostering the integration of commercial or industrial access technologies (i.e. Wi-Fi) into a single core network, jointly managing mobile devices as well as legacy or traditional devices. Thus, this paper focuses on enabling an Information-Theoretic Secure (ITS) framework for the non-3GPP access of 5G networks.

\section{Proposed system}
\label{sec:proposed-system}

This section presents a system to integrate quantum-based security in the untrusted non-3GPP access procedure of 5G networks. The system operates in a scenario where a UE on an untrusted non-3GPP network accesses the 5G Core Network via the N3IWF. Each device is located at a distinct site, interconnected by a direct QKD link. A local QKD module and a Key Management Server (KMS) serve the network at each site to provide applications with information-theoretically secure keys. Although the scenario typically assumes the N3IWF is located alongside other 5G Core Network Functions, this is not a strict requirement.

The UE and N3IWF interface with the QKD system through a standardized API that delivers QKD keys. The other Network Functions involved in the connection procedure are entirely agnostic of the QKD system’s presence.

Within this context, the proposed solution involves integrating QKD-generated keys into the IKEv2 handshake, securing the handshake and subsequently established IPsec Security Associations (SAs) between the UE and N3IWF.

\subsection{IKEv2 handshake with QKD}

\begin{figure}[h]
\centering
\includegraphics[width=1\columnwidth]{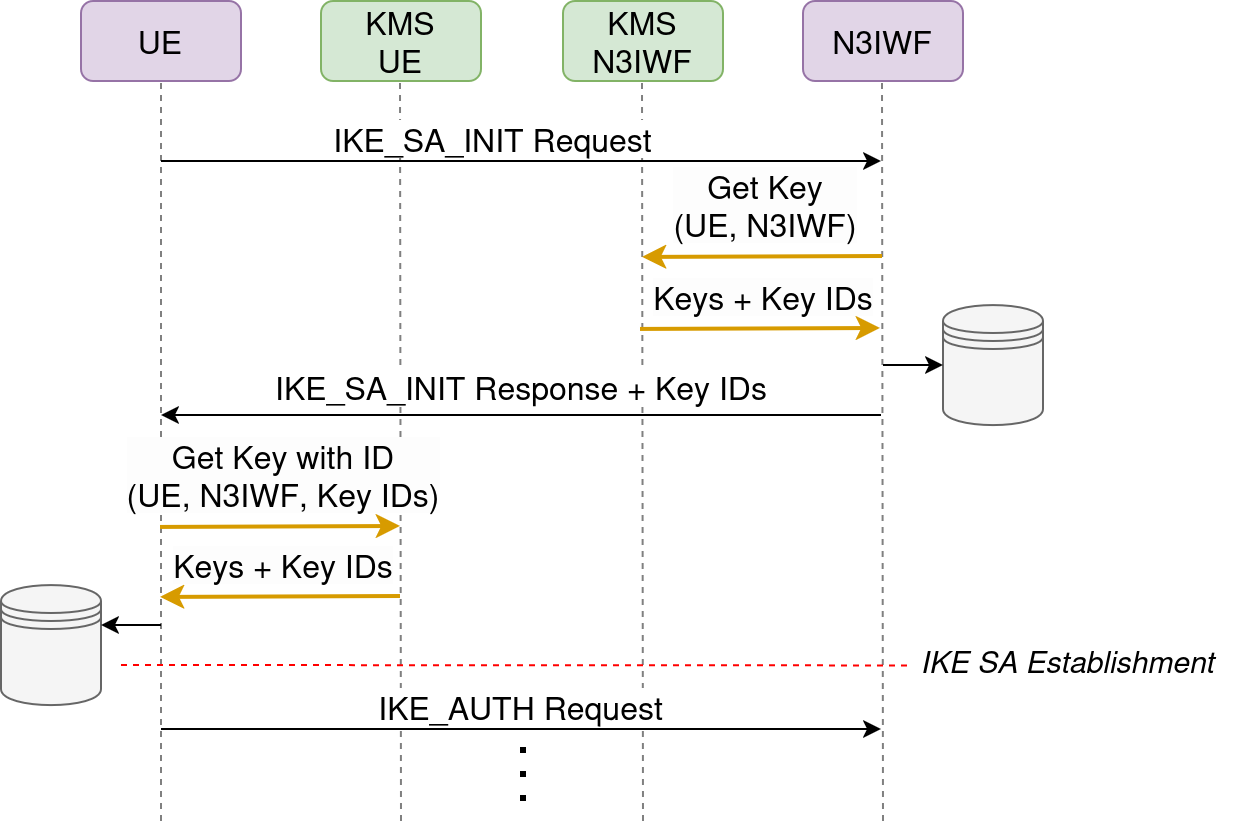}
\caption{Message exchange during the initial IKEv2 handshake among the 5G entities.}
\label{fig:ikev2}
\end{figure}

The integration of QKD into the untrusted non-3GPP connection establishment, depicted in Fig.~\ref{fig:ikev2}, occurs during the initial IKEv2 handshake between the UE and the N3IWF. The following steps describe the key retrieval phase, focusing on communication between the UE, the N3IWF, and their respective KMSs. This process is facilitated by the implementation of the ETSI GS QKD 014 API~\cite{ETSI014} in the involved KMSs.

\begin{enumerate}
    \item The UE initiates the IKEv2 handshake by sending an \texttt{IKE\_SA\_INIT request} message, including the Security Association proposal with the supported algorithms and parameters, as in the standard IKEv2. Notably, the proposal omits the Diffie-Hellman Group Transform and the Key Exchange material payloads, as the required keys will be obtained via QKD.
    \item Upon processing the request, the N3IWF retrieves the necessary keys from its associated KMS:
    \begin{itemize}
        \item The N3IWF directly requests all the keys required in the IKEv2 handshake via a single \texttt{Get key} request to its KMS, specifying the UE's identity.
        \item The KMS responds by providing the requested keys along with the Key Identifiers (Key IDs) assigned to those keys, so that the UE can later retrieve said keys.
    \end{itemize}
    \item Upon receiving the keys, the N3IWF stores them, creates the initial IKE SA, and sends all the Key IDs to the UE inside a Notify payload in the \texttt{IKE\_SA\_INIT response} message, along with its own Security Association proposal. As on the initiator side, the responder's proposal omits any payload related to the Diffie-Hellman exchange.
    \item The UE then processes the incoming \texttt{IKE\_SA\_INIT response}, extracts the Key IDs, and uses them to call the \texttt{Get key with ID} method from its KMS to obtain the keys. The UE stores the keys and then proceeds to create its IKE SA.
\end{enumerate}

Following the IKE\_SA\_INIT exchange, the N3IWF and the UE continue with the connection establishment by performing the untrusted non-3GPP authentication procedure in the IKE\_AUTH phase. Subsequently, both parties make use of the QKD keys stored during the initial phase to establish the Control Plane and User Plane Child SAs.

The amount of symmetric keying material required is determined by how keys are derived in standard IKEv2. For each IKE SA, four distinct keys are generated: two for encryption, and two for integrity protection, with one key per traffic direction in each case. In addition, every Child SA needs four keys: two for encryption and two for integrity protection. If Pre-Shared Key (PSK) authentication is used, one extra key is needed. As a result, for the NWu interface between a non-3GPP UE and the N3IWF, where deployments typically establish two Child SAs (separating control-plane and user-plane traffic), the total number of symmetric keys required is 13.

\section{Testbed}
\label{sec:testbed}

\begin{figure*}[t]
\centering
\includegraphics[width=0.8\linewidth]{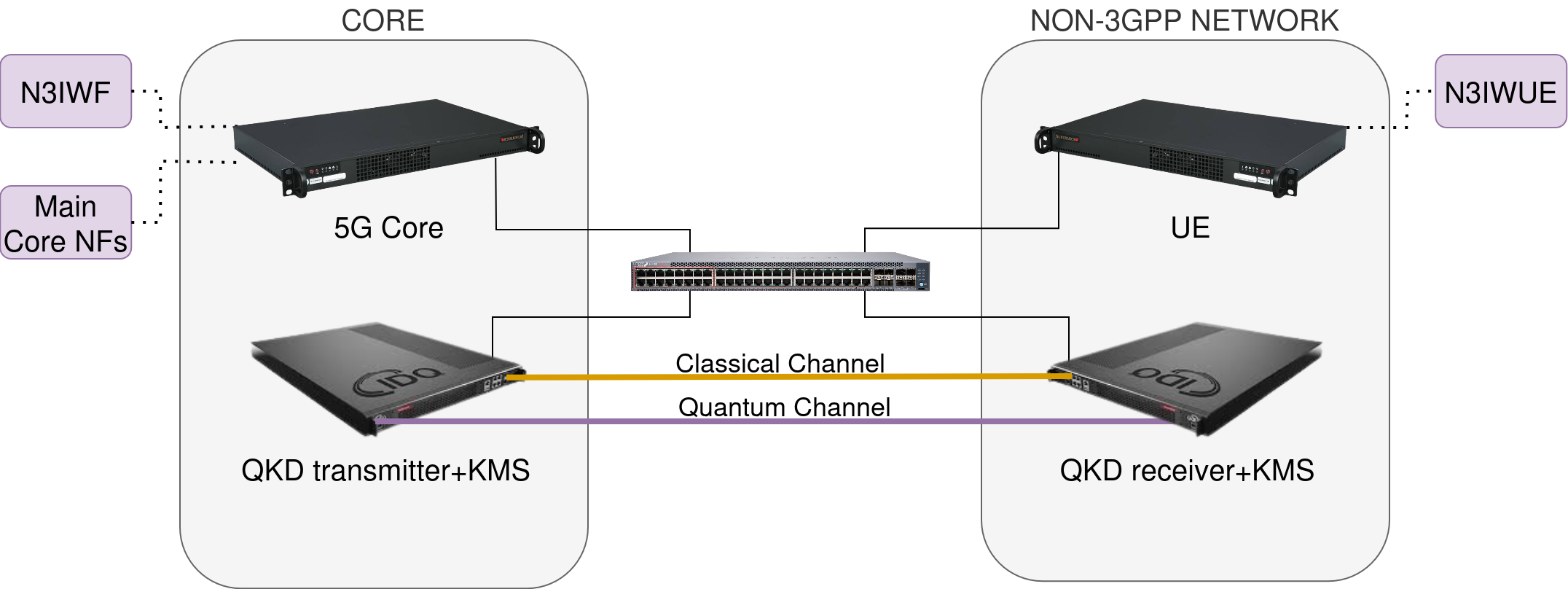}
\caption{Testbed for the experimental validation of the proposed system and the measurement of KPIs.}
\label{fig:testbed}
\end{figure*}

This section presents the testbed used to evaluate the proposed system. As depicted in Fig.~\ref{fig:testbed}, the testbed consists of two distinct network segments interconnected by a network switch. The first segment hosts the entire 5G Core, including the N3IWF. The second segment emulates an untrusted non-3GPP network, containing a UE that connects to the 5G Core via the N3IWF. The quantum network is distributed between both segments, with each containing a QKD device that acts as a network endpoint.

The quantum network is based on real-world prepare-and-measure discrete-variable QKD technology, implemented using a cutting-edge Clavis XGR system from ID Quantique~\cite{idquantiqueClavisSystem}. The system comprises a transmitter in the 5G Core location and a receiver in the untrusted non-3GPP access segment. Additionally, this system implements the ETSI GS QKD~014 API to deliver QKD keys to applications.

The 5G network functions are instantiated on two servers, one in each network segment. The server that hosts the 5G Core is a Supermicro SYS-E200-8D, with an Intel(R) Xeon(R) CPU D-1528 processor running at 1.90~GHz on an x86\_64 architecture with 6 cores (12 threads), 64~GB of RAM, 500~GB of local storage, and running Ubuntu 22.04. This server employs the free5GC (version 4.0.1)~\cite{free5gcFree5GC} open-source implementation to emulate the main core network functions. This setup uses a modified version of the N3IWF that integrates QKD into the default IKEv2 implementation. 

Similarly, the server that hosts the untrusted non-3GPP network segment is a Supermicro SYS-5028DTN4T, with an Intel(R) Xeon(R) D-1541 processor running at 2.10~GHz on an x86\_64 architecture with 8 cores (16 threads), 64 GB of RAM, 500 GB of local storage, and running Ubuntu 18.04. This server runs a QKD-enabled version of the N3IWUE application, which also includes modifications to its IKEv2 stack. Apart from the specified IKEv2 modifications for QKD integration, no further changes have been made to the original free5GC software components.


\subsection{Definition of KPIs}

Two different Key Performance Indicators (KPIs) are defined to assess the performance of the proposed solution.

\begin{enumerate}
    \item \textbf{Connection establishment time:} Duration of a connection establishment procedure between the UE and N3IWF. This interval starts with the initial IKE\_SA\_INIT message and concludes immediately before the connectivity tests performed by the N3IWUE application following PDU session establishment. 
    \item \textbf{Communication overhead (packet size):} The size of the packets transmitted during connection establishment. This parameter considers only the IKEv2 packets exchanged during the procedure, with the objective of evaluating the impact of QKD integration.
\end{enumerate}

\subsection{Testing methodology}

Several tests have been carried out to evaluate the proposal's feasibility and performance relative to the defined KPIs. The tests focus on the connection establishment procedure, analyzing the duration of the phases of the procedure to precisely evaluate the impact of QKD key usage in each. Furthermore, all test traffic was captured to analyze packet sizes and their payload data.

Three different configurations have been compared in the tests: 

\begin{itemize}
    \item The proposed system, that replaces conventional cryptographic methods (such as Diffie-Hellman key exchange and key derivation) with QKD keys.
    \item The standard IKEv2 operation with Diffie-Hellman key exchange and a PSK-based authentication system, implementing the 5G-specific version where keys are derived from subscriber secrets.
    \item The standard IKEv2 operation with Diffie-Hellman key exchange and a certificate-based authentication system.
\end{itemize}

It is important to note that non-QKD alternatives differ only in their peer authentication method, while both rely on Diffie-Hellman and PRF functions to derive keys for encryption and integrity checks.

These configurations were selected to provide a meaningful comparison of  the proposed solution, not only with the current standard-mandated method but also against a common alternative. Each test was executed 100 times to identify anomalies and obtain statistically significant results.

\section{Results and discussion}
\label{sec:results}

This section presents the measurement results along with an in-depth analysis. Regarding the first KPI, the connection establishment time, the measurements are divided into three main phases:

\begin{enumerate}
    \item \textbf{INIT phase:} IKE SA INIT message exchange, where all the cryptographic keys are negotiated and established.
    \item \textbf{AUTH phase:} IKE AUTH message exchange, including IKE authentication, EAP-5G authentication and control-plane SA creation.
    \item \textbf{Child SA phase:} CREATE CHILD SA message exchange, which includes the control-plane establishment, the user-plane SA creation and user-plane establishment.
\end{enumerate}

\begin{figure}
\includegraphics[scale=0.5]{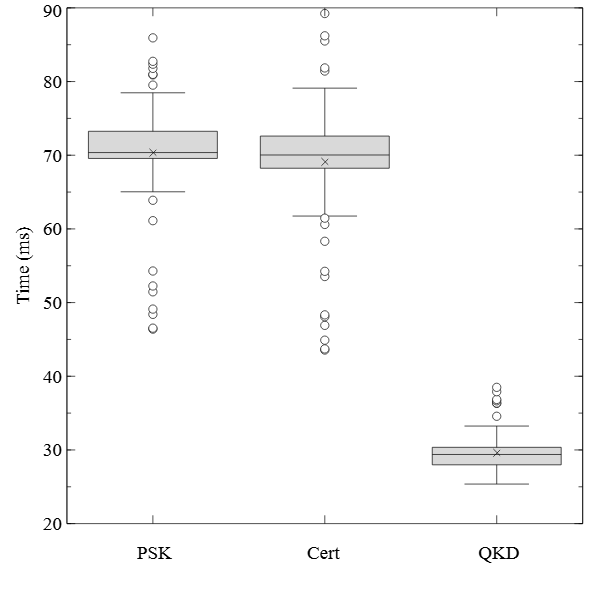}
\caption{Connection establishment time (INIT phase) for PSK-based, certificate-based and QKD-based systems.}
\label{fig:init}
\end{figure}

Additionally, as described in the methodology section, the tests have been performed for three different configurations: (1) D-H key exchange and PSK-based authentication, (2) D-H key exchange and certificate-based authentication, and (3) the proposed system based on QKD. Fig.~\ref{fig:init} depicts the required time for the first phase, which corresponds to the IKE INIT phase. The results are presented in boxplot format, where the box indicates the inter-quartile range (IQR), 25th to 75th percentiles, and the whiskers extend to 1.5 times the IQR. The mean value for each case is represented with an 'x' marker, and outliers with circles outside the whiskers.

Results show that the PSK-based system and the certificate-based system require 69.09 milliseconds and 70.34 milliseconds on average respectively, while the proposed system based on QKD requires 29.61 milliseconds on average. The drastic reduction of 40 milliseconds corresponds to the Diffie-Hellman exchange and consequent derivations that are executed in the first two cases, which are not necessary in the proposed system and are instead replaced by two REST queries (based on ETSI GS QKD 014 API) to the QKD equipments to retrieve the 15 necessary keys directly. The QKD queries are performed from the end node to the KMS, one on each side of the communication, and are placed within the same site to maintain security guarantees. In contrast, apart from the message exchange, the Diffie-Hellman exchange is composed by two exponential functions on each end. The impact of these functions has been measured in the testbed, requiring 8.81 milliseconds and 12.19 milliseconds in the case of the N3IWF server and the non-3GPP UE server respectively. As these functions are performed twice on each end, it results in 42 milliseconds, which is roughly the difference with respect to the QKD-based system. Thus, the reduction on the IKE INIT phase clearly corresponds to the replacement of the Diffie-Hellman exchange with \texttt{Get Key} and \texttt{Get Key with ID} queries to the KMS servers to retrieve the QKD keys.

\begin{figure}
\includegraphics[scale=0.5]{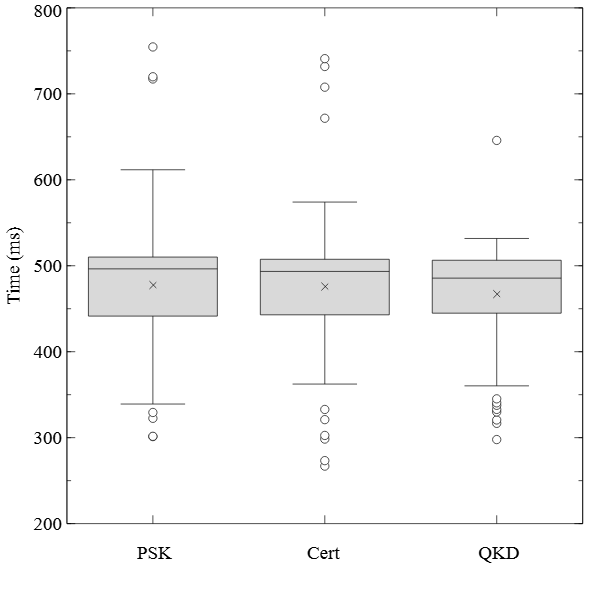}
\caption{Connection establishment time (AUTH phase) for PSK-based, certificate-based and QKD-based systems.}
\label{fig:auth}
\end{figure}

Fig.~\ref{fig:auth} depicts the required time for the second phase, which corresponds to the IKE AUTH message exchange that performs the authentication and generates the first Child SA for control-plane traffic. In this case, the PSK and certificate-based solutions require on average 475.81 and 477.51 milliseconds respectively, while the proposed system based on QKD requires 467.02 milliseconds. The time reduction for this phase is significantly lower, which is remarkable considering that the phase duration is on average 10 times higher than the INIT phase. In fact, the subprocesses replaced in this phase are PRFs implemented using hash functions, which are generally considered fast and time-efficient. However, as these functions are replaced by variable manipulation functions (i.e. read values, copy, etc.), which are even faster, the AUTH phase is reduced by 10 milliseconds in the proposed system.

\begin{figure}
\includegraphics[scale=0.5]{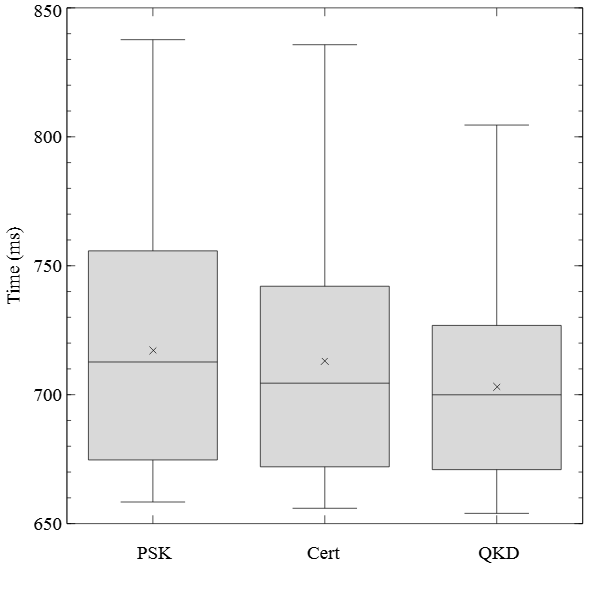}
\caption{Connection establishment time (Child SA phase) for PSK-based, certificate-based and QKD-based systems.}
\label{fig:childsa}
\end{figure}

Fig.~\ref{fig:childsa} shows the time required to complete the last phase, which involves the creation of the Child SA for the user-plane connections. This is the phase that requires the longest time to complete, with comparable relative results as the previous phase with 712.89 milliseconds and 717.15 milliseconds on average for PSK and certificate-based systems, respectively. Meanwhile, the proposed system based on QKD requires 703.03 milliseconds, improving the other results in 10 to 14 milliseconds. The reason behind this reduction is also similar to the previous phase, which results in this improvement. Table~\ref{tab:times} shows a summary of the results, including the mean and the standard deviation for each IKE phase and across the three analyzed configurations.

\input{table1}

Regarding the communication overhead, measured by the packet sizes, results can be seen in Table~\ref{tab:overhead}. The complete communication is performed with a series of message exchanges, which have different sizes depending on the authentication method used. In this context, the system that requires the most overhead is the certificate-based system, as it requires the transmission of the certificate in the \textit{IKE AUTH MID=01 I} message. Likewise, the QKD-based approach requires the transmission of the Key IDs from the N3IWF to the UE in the \textit{IKE SA INIT MID=00 R} message, which increases the packet size. However, the overall packet sizes from that point onward are similar or smaller than the PSK-based system, resulting in a lower total overhead for the proposed system. Nonetheless, it must be noted that the QKD-based system requires additional message exchanges between the N3IWF/UE and the KMS servers, although this communication is not directly part of the IPsec establishment and it stays within what is considered a local domain. Therefore, it can be said that the overhead introduced by IPsec in the link between the UE and the N3IWF is reduced with the proposed system.

\input{table2}

\section{Conclusions}
\label{sec:conclusions}

This paper proposes a system to integrate the QKD technology in the non-3GPP access to 5G networks leveraging IPsec protocol suite. The system replaces the regular Diffie-Hellman key exchange with PSK-based or certificate-based authentication mechanism of the IKE protocol with a QKD-based mechanism, obtaining keys through standardized APIs from the KMS servers in the initial phase of the IKE message exchange. The proposed system was implemented in a hardware-based testbed with commercial QKD equipment, integrating it in a virtualized and modularized 5G network. Results demonstrate a performance improvement regarding the time required to establish the IPsec connection, as well as reducing the overhead of the communication by reducing the packet sizes. In fact, the time required to complete the authentication and IPsec SA establishment is 4.62\% faster than the PSK-based system and 5.17\% faster than the certificate-based system. However, even though enhancing the performance of the overall non-3GPP access to the 5G network, the main benefit of the proposed system is the adoption of 5G security to the quantum era. 

The proposed system fosters the implementation of quantum-safe IPsec within 5G networks, which is necessary for the protection of current communications against the threat posed by the advent of quantum computing. In this context, additional efforts could be directed toward adapting 5G and 6G networks for the quantum era. QKD and similar quantum-aware technologies can be utilized to enhance other aspects of communication, as the security protocols identified in the specifications, such as TLS, exhibit similar functions and interactions. Furthermore, the quantum security of the 3GPP access for UE should also be evaluated. In this regard, alternative technologies such as Post-Quantum Cryptography (PQC) may be more suitable than QKD, since securing the KMS access from wireless devices outside the QKD domain can be challenging and may require additional techniques. This could potentially impact performance, increase complexity, and expand the attack surface. Consequently, substantial progress remains to be made in this field, given the critical importance of secure communications in the modern era.

\section*{Acknowledgment}

This work was supported in part by the Spanish Ministry of Science and Innovation in the project EnablIng Native-AI Secure deterministic 6G networks for hyPer-connected envIRonmEnts (6G-INSPIRE) (PID2022-137329OB-C44), and in collaboration with the National Hub of Excellence in Quantum Communications, funded by the Ministry of Digital Transition and Public Administration with European Union funds – Next Generation EU, under Component 16.R1 of the Recovery, Transformation, and Resilience Plan.

\bibliographystyle{IEEEtran}
\bibliography{refs}

\vspace{12pt}

\end{document}

%% file: table1.tex
\begin{table}[]
\caption{Required time for the complete connection establishment for PSK-based, certificate-based and QKD-based systems.}
\label{tab:times}
\centering
\resizebox{\columnwidth}{!}{%

\begin{tabular}{|c|c|c|c|}
\hline
                        & \textbf{PSK-based}                                                  & \textbf{Certificate-based}                                          & \textbf{QKD-based}                                                  \\ \hline
\textbf{INIT phase}     & \begin{tabular}[c]{@{}c@{}}M: 69.09 ms\\ SD: 7.98 ms\end{tabular}   & \begin{tabular}[c]{@{}c@{}}M: 70.34 ms\\ SD: 6.39 ms\end{tabular}   & \begin{tabular}[c]{@{}c@{}}M: 29.61 ms\\ SD: 2.53 ms\end{tabular}   \\ \hline
\textbf{AUTH phase}     & \begin{tabular}[c]{@{}c@{}}M: 475.81 ms\\ SD: 78.49 ms\end{tabular} & \begin{tabular}[c]{@{}c@{}}M: 477.51 ms\\ SD: 74.30 ms\end{tabular} & \begin{tabular}[c]{@{}c@{}}M: 467.02 ms\\ SD: 58.32 ms\end{tabular} \\ \hline
\textbf{Child SA phase} & \begin{tabular}[c]{@{}c@{}}M: 712.89 ms\\ SD: 43.36 ms\end{tabular} & \begin{tabular}[c]{@{}c@{}}M: 717.15 ms\\ SD: 45.15 ms\end{tabular} & \begin{tabular}[c]{@{}c@{}}M: 703.03 ms\\ SD: 36.18 ms\end{tabular} \\ \hline
\textbf{TOTAL}          & \textbf{M: 1257.79 ms}                                              & \textbf{M: 1265 ms}                                                 & \textbf{M: 1199.66 ms}                                              \\ \hline
\end{tabular}
}
\end{table}

%% file: table2.tex
\begin{table}[]
\centering
\caption{Communication overhead (packet sizes) for PSK-based, certificate-based and QKD-based systems, measured in bytes.}
\label{tab:overhead}
\begin{tabular}{|c|c|c|c|}
\hline
                                & \textbf{PSK}  & \textbf{Certificate} & \textbf{QKD}  \\ \hline
\textbf{IKE\_SA\_INIT MID=00 I} & 698           & 698                  & 428           \\ \hline
\textbf{IKE\_SA\_INIT MID=00 R} & 698           & 698                  & 1099          \\ \hline
\textbf{IKE\_AUTH MID=01 I}     & 214           & 1494                 & 214           \\ \hline
\textbf{IKE\_AUTH MID=01 R}     & 1446          & 1446                 & 1446          \\ \hline
\textbf{IKE\_AUTH MID=02 I}     & 182           & 182                  & 182           \\ \hline
\textbf{IKE\_AUTH MID=02 R}     & 166           & 166                  & 166           \\ \hline
\textbf{IKE\_AUTH MID=03 I}     & 150           & 150                  & 150           \\ \hline
\textbf{IKE\_AUTH MID=03 R}     & 150           & 150                  & 150           \\ \hline
\textbf{IKE\_AUTH MID=04 I}     & 182           & 182                  & 182           \\ \hline
\textbf{IKE\_AUTH MID=04 R}     & 118           & 118                  & 118           \\ \hline
\textbf{IKE\_AUTH MID=05 I}     & 150           & 118                  & 150           \\ \hline
\textbf{IKE\_AUTH MID=05 R}     & 278           & 246                  & 278           \\ \hline
\textbf{CHILD\_SA MID=05 R}     & 286           & 486                  & 230           \\ \hline
\textbf{CHILD\_SA MID=05 R}     & 470           & 470                  & 198           \\ \hline
\textbf{TOTAL}                  & \textbf{5388} & \textbf{6604}        & \textbf{4991} \\ \hline
\end{tabular}
\end{table}